# Demonstrating REACT: a Real-time Educational AI-powered Classroom Tool

Ajay Kulkarni
George Mason University
akulkar8@gmu.edu

Olga Gkountouna
George Mason University
ogkounto@gmu.edu

## ABSTRACT

We present a demonstration of REACT, a new Real-time Educational AI-powered Classroom Tool that employs EDM techniques for supporting the decision-making process of educators. REACT is a data-driven tool with a user-friendly graphical interface. It analyzes students' performance data and provides context-based alerts as well as recommendations to educators for course planning. Furthermore, it incorporates model-agnostic explanations for bringing explainability and interpretability in the process of decision making. This paper demonstrates a use case scenario of our proposed tool using a real-world data set, and presents the design of its architecture and user-interface. This demonstration focuses on the agglomerative clustering of students based on their performance (i.e., incorrect responses and hints used) during an in-class activity. This formation of clusters of students with similar strengths and weaknesses may help educators to improve their course planning by identifying at-risk students, forming study groups, or encouraging tutoring between students of different strengths.

## 1. INTRODUCTION

Instructors play a crucial role in educational institutions, where one of their main responsibilities is effective high-quality teaching. To do so they must stay updated with students' responses, efforts, and outcomes, in order to provide timely feedback to promote students' improvement [9, 29]. One of the ways this can be achieved is by clustering students into groups based on various characteristics such as their learning style preferences, academic performance, behavioral interaction, etc., which can be utilized to explore collaborative learning opportunities and identify at-risk students at an early stage [3].

Thus, this creates a need for tools that will empower instructors to achieve these objectives in the classroom. To this end, the fields of Educational Data Mining (EDM) and Learning Analytics (LA) have emerged with the goal to understand how educational data can benefit the science of learning [7]. One of the ways to promote this understanding is to use AI-powered real-time visualizations. These visual displays summarize large amounts of data in a meaningful way. This is important for humans' sense-making and decision-making [35] as it helps human cognition [12]. An example of these visual displays are dashboards which may contain various data indicators [20].

Furthermore, applications of Artificial Intelligence (AI) in the domain of education for predicting student performance, detecting undesirable student behavior, or providing feedback for supporting instructors and students, are becoming more common [8]. This creates a need for incorporating interpretability, explainability, and, ultimately, trustworthiness in AI for supporting human teaching and learning [39]. The simplest way to include explainability in AI is by using model-agnostic explanations that consist of textual and visual explanations [4]. Interpretability can be achieved by including humans in the process of decision making (HitAI) i.e., decision power is given to the specialized professionals who utilize machines/tools as advisors [41].

This paper presents a demonstration of REACT, a Realtime Educational AI-powered Classroom Tool, which utilizes the principles of HitAI and model-agnostic explanations to support educators in their decision-making. REACT clusters students based on their responses during in-class activities, and provides context-based recommendations for course planning. It also provides personalized feedback about individual students. REACT is a real-time data-driven decision support tool that incorporates explainability, interpretability, and portability. It presents different indicators of students, their learning processes, learning contexts, and recommendations for increasing efficiency in course management. Based on the Learning Analytics Process model [38], REACT may directly help educators in awareness, reflection, and sense-making while it can indirectly create impact and motivate to take actions. This functionality can support educators in making decisions concerning their course planning and instructional goals setting, by consistently monitoring students' activities (tests, quizzes, and exercises) to inspect the their learning process.

The remainder of this paper is structured as follows. Section 2 presents the background of EDM and how clustering can be useful in this context. Section 3 describes the architecture of our tool and explains how clustering is utilized in REACT. Section 4 presents the details on the design of the user interface and details of the demonstration. Finally, Section 5 concludes the paper discussing directions of future research.

## 2. BACKGROUND

EDM enhances the decision-making of teachers, students, and educational institutes by utilizing data mining techniques in an educational context [31]. A variety of methods, including cluster analysis, outlier detection, text mining, recommendation systems, and visualizations can be applied in the EDM domain [32]. *Cluster analysis* or *clustering* is the most well known unsupervised machine learning task [24]. In the context of EDM, identifying meaningful clusters of students can be useful in understanding their learning behavior [14, 13, 10, 5, 26, 22]. Clustering consists of four steps [40]: (1.) *Feature extraction and selection* – Relevant features are selected from the data and transformed into an appropriate format. (2.) *Algorithm design* – A suitable clustering algorithm and (dis)similarity measure are selected. (3.) *Evaluation* – Different clustering results can be evaluated using different metrics such as external, internal and/or relative indices. (4.) *Explanation* - The main purpose of clustering is to generate knowledge, useful for decision-making. This is conveyed to the user in different ways such as visualizations, textual feedback, or statistical metrics.

Hierarchical clustering organizes the data points in a taxonomy tree of clusters and sub-clusters [37]. Thus, it is suitable for detecting clusters of arbitrary shape, type and hierarchical relationships [40]. Hierarchical clustering has been shown to provide good results for small datasets [1], which is useful for typical class sizes. Furthermore, the entire clustering process can be visualized by plotting a *dendrogram*, which shows the cluster-subcluster relationships, similarity between clusters, and the order in which they are merged [37]. This results in an informative visualization of the data clustering structures [40], fulfilling the goal of explainability. There are two basic approaches to Hierarchical clustering – agglomerative and divisive [25]. The agglomerative hierarchical clustering is a bottom-up approach that starts with each points being an individual cluster, and merges the closest pair of clusters at each step. The divisive method is top-down approach that starts with points being in one large cluster and progressively divides them. This is computationally expensive [15] and not commonly used [40]. Thus, agglomerative hierarchical clustering makes a suitable choice for implementing cluster analysis on REACT.

## 3. REACT ARCHITECTURE

REACT is developed using the `R Shiny` framework which incorporates the principles of reactive programming [6] that are suitable for interactive applications. REACT is portable in a sense that it can be connected to any Learning Management System (LMS), like Moodle or Blackboard, as well as different database management systems, including MySQL, Oracle, Salesforce, etc. This can be achieved by using dif-ferent packages such as DBI[1] (for databases), bRush[2] and rcanvas[3] (for the Canvas LMS) which are available in R. Additionally, many other LMSs other REST APIs which can be connected with REACT using httr[4] and jsonlite[5] packages.

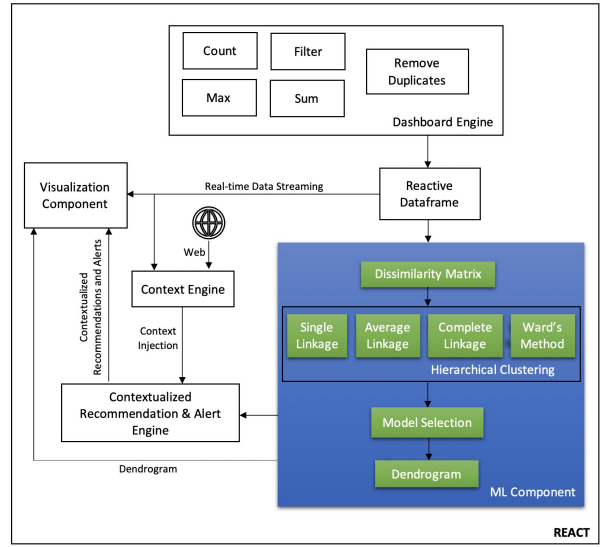

Figure 1: Architecture of REACT

The architecture of REACT is motivated from RAED [23] and it is shown in Figure 1. It consists of five main components: the Dashboard Engine, the Machine Learning (ML) Component, the Context Engine, the Contextualized Recommendation & Alert Engine, and the Visualization Component. Due to space limitations, we focus on the component that implements clustering as an EDM technique, i.e., the ML component.

The ML component receives input from a reactive data frame that contains the input features of each student and initiates the clustering process by first calculating all the pairwise distances (i.e., dissimilarities) of students. We use the Gower distance [18] as the dissimilarity metric for the clustering, as it can be applied to mixed data (i.e., a mix of numerical and categorical variables) in general [33]. However, for the purposes of this demonstration, we use as input features of each student the numbers of incorrect responses and the number of hints used per learning concept. For these numerical features, Gower uses Manhattan distance to calculate dissimilarity. The Dissimilarity Matrix sub-component calculates the pairwise distances between all *n* observations (i.e., students) in the data set organized in an n*n matrix, using the `daisy()R` function. This dissimilarity matrix then becomes the input of the Hierarchical Clustering sub-component.

The Hierarchical Clustering sub-component trains four different hierarchical clustering models using the same dissim-ilarity matrix. These models are based on four different linkage methods: Single linkage, Average linkage, Complete linkage, and Ward's method. The R function `agnes()` is used for building these models and computing their agglomerative coefficients.

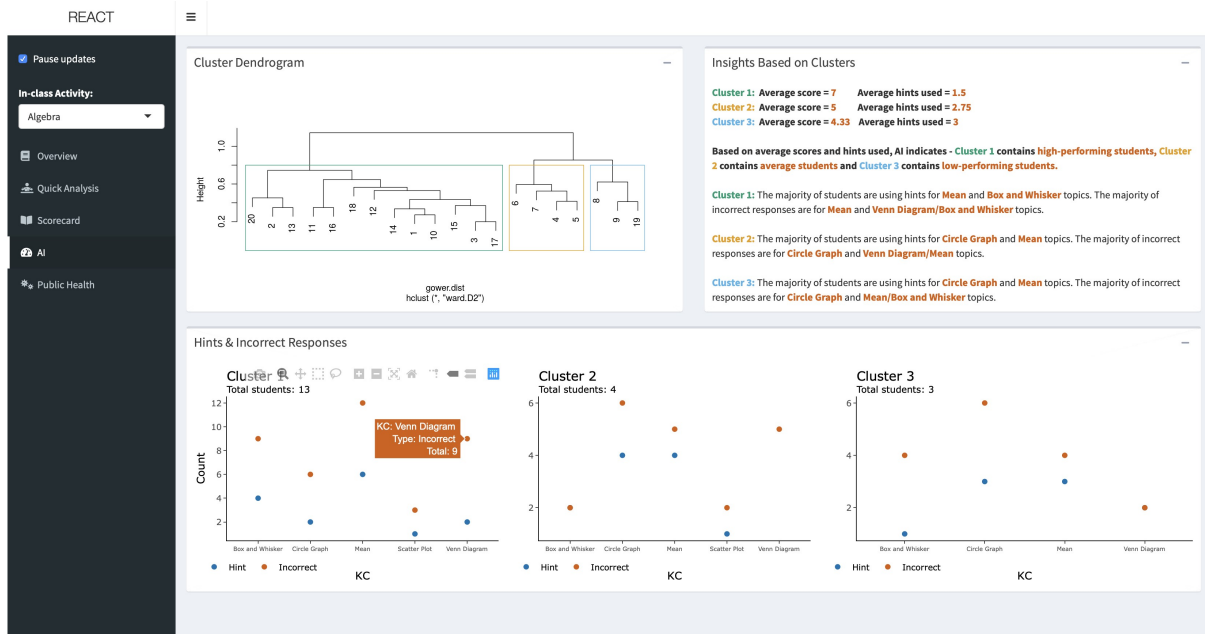

Figure 2: The AI tab provides real-time insights of clustering to instructors with visual explanation (dendrogram - top left) and textual template-based recommendations (top right)

The Model Selection sub-component ensures robustness and acts as an internal index for evaluations. It compares the four clustering results based on their agglomerative coefficient. Their values lie between 0 to 1, and describe the strength of the corresponding clustering structure [21]. This sub-component selects the model with the highest agglomerative coefficient.

The Dendrogram sub-component creates a visualization of the hierarchy of clusters and sub-clusters that are the result of the selected model. This visualized hierarchy is called a *dendrogram*. The dendrogram provides a diagrammatic representation of the hierarchical cluster analysis. It can help to understand the clustering process which may help to incorporate explainability. An example of a dendrogram is shown in Figure 2 and discussed in the next section.

## 4. DESIGN AND DEMO

A dashboard can be defined as "an easy to read, often single page, real-time user interface, showing a graphical presentation of the current status (snapshot) and historical trends of an organization's key performance indicators (KPIs) to enable instantaneous and informed decisions to be made at a glance" [11]. It is also common for decision-makers to use KPIs for understanding the performance or the deviation from the set target at a glance [28]. Thus, the user-interface of REACT is designed as an interactive dashboard, displaying KPIs to help teachers monitor and understand their student's learning performance. These KPIs include the minimum, maximum, median and mean scores of the class, as well as the number of students who have completed all the questions of the in-class activity thus far.

### 4.1 Design Elements

A dashboard's visual attraction significantly affects its perceived usefulness and its potential to bring change in users' behavior [27]. The design choices of REACT were made with this consideration in mind. The selection of visualizations is based on the chart suggestions provided by Abela [2] and a review provided by Schwendimann et al. [34]. Further, its color palettes were selected so that it is colorblindfriendly [17]. REACT contains interactive visualizations and tables. Interactive applications need to ensure that they are easy to learn, and effective as well as enjoyable to use [30]. To ensure this, we aimed to follow the 'golden rules' of interface design proposed by Shneiderman et al. [36] – strive for consistency, permit easy reversal of actions, keep users in control, and reduce their short-term memory load.

The user interface of REACT currently has five tabs: *Overview* – presents the KPIs, an interactive plot for monitoring students' performance, and textual alerts & recommendations for the instructors.

*Quick Analysis* – presents an interactive plot that monitors the progress of students in real time, and bar charts that count the incorrect responses, and hints used for each KC.

*Scorecard* – displays a histogram of the score distribution, and a dynamic table with students' information and scores, both updated in real-time.

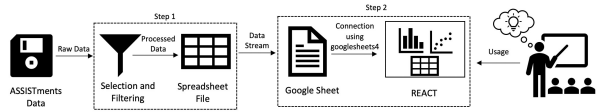

Figure 3: Creating a real-time demo of REACT

*AI* – provides insights of the cluster analysis and textual template-based recommendations. Figure 2 shows a screenshot of this tab. An instructor may use this tab to see on the dendrogram (top left) how different groups (i.e., clusters) of students are formed, based on their performance in in-class activities. Textual explanations about each of these groups are provided on the top right. Finally, to enhance interpretability, each of the different clusters is also visually explored at the bottom of the tab. The counts of incorrect answers and the counts of hints used are displayed per Knowledge Component (KC) for each group of students.

*Public Health* – provides context based on the COVID-19 outbreak. It displays infection rates in the surrounding counties and counts of students who may live in high risk areas, to inform educators, who may opt to transition online.

## 4.2 Demo

Holstein et al. [19] note the importance of using real-world datasets to understand the behavior of LA tools. We use the 2009-2010 Skill-builder ASSISTments data set [16]. The raw data consists of more than 100,000 rows representing details of 4217 students and 111 Knowledge Components (KCs). To achieve the objective of this demonstration, we randomly selected a sample of 20 students. Due to privacy, this data set includes only pseudo-ids. In real-word uses of REACT, authenticated instructors will be able to see students' names, as memorising their ids would be troublesome. Our approach to create a demonstration of REACT is shown in Figure 3 and can be summarized in the following steps:

- Step 1 (Filter): We selected 20 students and two questions from five KCs from the topic of statistics (Mean, Circle Graph, Venn Diagram, Box and Whisker Plot, and Scatter Plot). This filtered data set is first stored in a spreadsheet on a local hard disk.
- Step 2 (Stream): The filtered data from *Step 1* are then streamed on a Google sheet that acts as a database for this demonstration. It is connected to REACT using the `googlesheets4`[6] package.
- Step 3 (Use): REACT receives live updates from the streaming data. These concern hints that each student uses during the simulated in-class activity, as well as if they provided a correct or incorrect response to each question, as time progresses. These data are processed on the fly and used to update the visualisations, alerts, and recommendations displayed on the user interface.

A live version of REACT[7] is deployed using the `Shiny Server` and it can be accessed using a web browser on any desktop, laptop, tablet, or smartphone.

[6] `https://googlesheets4.tidyverse.org`
[7] `https://tinyurl.com/y7cbbbej`

## 5. CONCLUSIONS AND FUTURE WORK

We presented REACT, a data-driven, visual, decision-support tool that incorporates model-agnostic explanations. This paper provides details on demonstrating a use-case scenario by utilizing the ASSISTments dataset. Our next step is to evaluate our proposed tool with the help of domain experts, using a combined approach of think-aloud testing and questionnaires. This approach will help us to understand the usability and user experience while interacting with REACT. The results from this combined approach can help us to identify directions of improvement in the interface design and to propose the addition of new features. In the future, we aim to answer *how the integration of AI and visualizations in real-time can impact the instructors' decisionmaking process, and to what extent they do trust it.* The answers to these questions will play a crucial role in making REACT a deployable tool that can enhance data-driven decision-making in education.